\documentclass[aps,pra,showpacs]{revtex4}

\usepackage{graphicx}

\begin{document}

\title{On the estimation of interaction parameters in weak measurements}

\author{Holger F. Hofmann}
\email{hofmann@hiroshima-u.ac.jp}
\affiliation{
Graduate School of Advanced Sciences of Matter, Hiroshima University,
Kagamiyama 1-3-1, Higashi Hiroshima 739-8530, Japan}
\affiliation{JST,Crest, Sanbancho 5, Chiyoda-ku, Tokyo 102-0075, Japan
}

\begin{abstract}
Since weak measurements are known to produce measurement values that can be much larger than the maximal eigenvalues of the measured observable, it is an interesting question how this enhancement of the measurement signal relates to the sensitivity of quantum measurements as investigated in the field of quantum metrology. In this presentation, it is pointed out that the estimation of a small interaction parameter using weak measurements actually corresponds to standard quantum metrology, where the logarithmic derivatives of the final measurement probabilities are proportional to the corresponding weak values. The analysis of the general weak measurement formalism then shows that extreme weak values do not improve the sensitivity. Instead, all final measurements with real weak values have the same sensitivity as a final measurement of the eigenvalues. This result supports the view that real weak values can be interpreted as precise, zero uncertainty estimates of a quantum observable, despite their deviation from the eigenvalues of the corresponding operator.
\end{abstract}

\pacs{
03.65.Ta, 
03.67.-a 
42.50.St 
}

\maketitle


Both quantum metrology and weak measurements are based on small perturbations of the initial quantum state caused by a weak interaction with the quantum system. In quantum metrology, the goal is to estimate an unknown parameter such as a small phase shift $\phi$ from its effects on the outcomes of a final measurement $f$. In conventional weak measurements, the interaction parameter is fixed and the effects of post-selecting a final measurement outcome $f$ on the distribution of the weak measurement outcomes is considered. However, it has long been understood that the relation between the observed pointer shift of a weak measurement and the known weak value of a specific post-selected measurement can also be used to determine the value of an unknown measurement interaction parameter \cite{Hos08, Dix09}. Surprisingly, there seems to be a lack of work clarifying the relation between the outcomes of weak measurements and the sensitivity of the corresponding parameter estimation. Although the possible role of weak measurements in optical interferometry has recently been addressed \cite{Bru10}, the focus there has been more on the specific technical realizations of the measurement than on the conceptual correspondence between weak measurements and quantum metrology. In the following, I will try to bridge the gap between parameter estimation and weak measurements by formulating the quantum theory of a general weak measurement as a parameter estimation problem.  
Weak measurements can then be seen as a natural part of quantum metrology, providing new insights into the fundamental properties of quantum measurements and quantum statistics.


In general, weak measurements can be formulated in terms of measurement operators that are essentially independent of the technical details of system-pointer coupling \cite{Hof10a}. Every weak measurement outcome $m$ then corresponds to a self-adjoint measurement operator $\hat{E}_m$ that represents the quantum statistics of the measurement. Since we are considering weak measurements, the output probabilities $w_m$ are only slightly modified by the target observable $\hat{A}$. The magnitude of this modification depends on the strength of the measurement interaction. To formulate weak measurements as a parameter estimation problem, this coupling strength should be represented by an unknown interaction parameter $\epsilon$. The measurement operators can then be expressed as
\begin{equation}
\hat{E}_m = \sqrt{w_m} \left(\hat{1}+\epsilon \kappa_m \hat{A}\right),
\label{eq:measurement}
\end{equation}
where $\kappa_m$ represents the correlation between the outcome and the effects of the coupling for the specific type of measurement interaction. A convenient normalization of this correlation factor is
\begin{equation}
\sum_m w_m \kappa_m^2 = 1.
\label{eq:norm}
\end{equation}
Eq.(\ref{eq:measurement}) defines the parameter estimation problem in terms of an unknown parameter in a quantum mechanical operator acting on an input state. This is essentially equivalent to the phase shift problem, where the parameter modifies the unitary transformation acting on the quantum state. In fact, the operators $\hat{E}_m$ can be transformed into a random unitary transformation by replacing the real interaction parameter with an imaginary phase parameter. As pointed out in \cite{Hof10b}, this means that the estimation of a phase is equal to the estimation of the interaction parameter in a measurement of an imaginary weak value.

Ultimately, the parameter estimation procedure must be based on a final measurement represented by a set of orthogonal states $\{\mid f \rangle\}$. The joint probability of obtaining a weak result $m$ and a final 
measurement result $f$ is then given by 
\begin{equation}
p(m,f)= w_m | \langle f \mid \psi \rangle|^2 \left(1+ 2 \epsilon \kappa_m \mbox{Re}\left(\frac{
\langle f \mid \hat{A} \mid \psi \rangle}{\langle f \mid \psi \rangle}\right)\right).
\label{eq:weakvalue}
\end{equation}
The dependence of this probability on the interaction parameter $\epsilon$ is given by the real part of the weak value of $\hat{A}$ for the initial state $\mid \psi \rangle$ and a post-selection of $\mid f \rangle$. Specifically, the logarithmic derivatives are proportional to the real weak values, 
\begin{equation}
\frac{\partial}{\partial \epsilon} \ln \left( p(m,f) \right)|_{\epsilon=0}
= 2 \kappa_m \mbox{Re}\left[\frac{
\langle f \mid \hat{A} \mid \psi \rangle}{\langle f \mid \psi \rangle}\right].
\end{equation}
Thus the weak values of a final measurement $\mid f \rangle$ define the relevant statistics for an estimation of the interaction strength. 


It is well known that the sensitivity of a parameter estimation problem is determined by the Fisher information, which is the inverse of the squared estimation error for an optimal measurement strategy. For the estimate of the interaction parameter $\epsilon$ described here, this sensitivity is given by
\begin{eqnarray}
\frac{1}{\delta \epsilon^2} &=&
\sum_{m,f} p(m,f) \left(\frac{\partial}{\partial \epsilon} \ln \left( p(m,f) \right)|_{\epsilon=0}\right)^2
\nonumber \\
&=& 4 \sum_f p(f) \; \mbox{Re}\left[\frac{
\langle f \mid \hat{A} \mid \psi \rangle}{\langle f \mid \psi \rangle}\right]^2,
\label{eq:weakFisher}
\end{eqnarray}
where the joint probability at $\epsilon=0$ is given by $p(m,f)=w_m p(f)$, so that the sum over $m$ can be determined using the normalization in eq.(\ref{eq:norm}). The optimal sensitivity is obtained for final measurements with completely real weak values,
\begin{equation}
\mbox{Im}\left[\frac{
\langle f \mid \hat{A} \mid \psi \rangle}{\langle f \mid \psi \rangle}\right]=0.
\end{equation}
For all measurements that fulfill this condition, $p(f)=|\langle f \mid \psi \rangle|^2$ cancels out the division by the corresponding probability amplitude $\langle f \mid \psi \rangle$ in the weak values, and the summation over $f$ converts into a general Hilbert space trace. The maximal sensitivity of the interaction parameter estimate is then given by four times the expectation value of the squared operator observable $\hat{A}$ in the initial state,
\begin{equation}
\frac{1}{\delta \epsilon^2} = \langle \psi \mid \hat{A}^2 \mid \psi \rangle.
\end{equation}
The most simple way of obtaining this sensitivity is to perform a final measurement of the eigenstates of $\hat{A}$ in the output. The weak values are then equal to the eigenvalues of $\hat{A}$ and the interaction parameter is determined from the pointer shifts obtained for each eigenvalue. This procedure can be understood classically as a precise determination of the unknown parameter $\hat{A}$ in the system-meter coupling. Since any uncertainty in $\hat{A}$ would add to the uncertainty of the estimate, it seems to be best to perform a precise measurement of $\hat{A}$. However, the same sensitivity can be achieved by any final measurement $\{\mid f \rangle\}$, as long as all of the weak values are real. It is therefore possible to vary the measurement strategy without losing sensitivity. In fact, it can be seen from eq.(\ref{eq:weakFisher}) that high weak values do contribute more to the Fisher information. However, this is compensated by their low probability $p(f)$. Effectively, it is possible to ``shunt'' the phase sensitivity into just a few measurement results. In the extreme case, it is possible to define a final measurement so that all weak values except one are zero, so that the parameter estimate can be derived from the frequency of only a single result. 

There are two important conclusions that can be drawn from the observation that the same maximal sensitivity can be obtained from any set of real weak values, whether they correspond to the eigenvalues or not. On the fundamental side, it is remarkable that the use of weak values in place of the unknown parameter $\hat{A}$ that determines the effects of the interaction on the measurement result $m$ does not add any noise compared to the use of the eigenvalues determined in a final von Neumann measurement. This result supports the viewpoint that real weak values represent zero uncertainty estimates of the observable $\hat{A}$ as expressed in the works of Hall \cite{Hal04} and Johansen \cite{Joh04}, based on the theories of Ozawa as presented in his price lecture at QCMC 2010 \cite{Oza03}. On the more practical side, it is significant that one should distinguish the improved resolution represented by the enhancement of the weak values from the sensitivity, similar to the same distinction recently made between resolution and sensitivity in the context of phase estimation from interference fringes \cite{Res07}. The enhancement of the weak values is still a powerful tool for obtaining conclusive data without the need for a thorough statistical analysis. Combined with the fact that the weak values do seem to provide a valid representation of the observable effects of physical properties, this distinction should help to motivate and guide further research into the intriguing statistics of quantum measurements.

\vspace{0.5cm}

Part of this work has been supported by the Grant-in-Aid program of the Japanese Society for the Promotion of Science, JSPS. I would also like to express my thanks to Akio Hosoya, who motivated me to take a closer look at weak measurements.

\end{document}